\documentclass[aps,prd,twocolumn,superscriptaddress]{revtex4}
\usepackage{epsfig,epsf}
\usepackage{amsmath}
\usepackage{amsthm}
\usepackage{amsfonts}
\usepackage{amssymb}
\usepackage{dsfont}
\usepackage{multirow}
\usepackage{appendix}
\usepackage{slashed}
\usepackage[active]{srcltx}
\usepackage{psfrag}

\begin{document}

\title{Newly observed resonance $ X(4685) $: diquark-antidiquark picture }
\author{A. T\"{u}rkan}
\affiliation{\"{O}zye\u{g}in University, Department of Natural and
Mathematical Sciences, \c{C}ekmek\"{o}y,Istanbul,Turkey}
\author{J.Y.~S\"ung\"u}
\affiliation{Department of Physics, Kocaeli University, 41001
Izmit, Turkey}\author{E. Veli Veliev}
\affiliation{Department of Physics, Kocaeli University, 41001 Izmit, Turkey}

\begin{abstract}
In this work, the mass and pole residue of
$ X(4685) $ state with spin-parity $J^P=1^+$ are computed by
the QCD sum rule approach up to operator dimension seven based on the diquark-antidiquark configuration.
For its mass we get $ m_{X_{cs}}= 4607^{+36}_{-22} $ MeV and pole residue $ \lambda_{X_{cs}}= 6.19^{+0.32}_{-0.24}\times 10^{-2}~\mathrm{GeV^5}$, which may be checked via other nonperturbative approaches as well as future experiments.  As a by product, the mass of the hidden-bottom partner state of the
$ X(4685) $ is extracted to be both around $ m_{X_{bs}}= (10604-10924) $ MeV and $ \lambda_{X_{bs}}= (42.2-53.7)\times 10^{-2}~\mathrm{GeV^5}$, which can be searched in the $\Upsilon \phi $ invariant mass distribution.

\end{abstract}

\maketitle

\section{Introduction}

As the universe cooled after the Big Bang, the building blocks of matter $-$ the quarks and electrons of which we are all made started to form. But, we do not have complete information 
on dynamics of quark and gluon creating hadrons, which can be better understood by searching exotic matter beyond the traditional quark model. Many unconventional multi-quark structures have been detected a plethora charmonium-/bottomonium-like XYZ states \cite{Brambilla:2019esw,Ali:2017jda,Liu:2019zoy,Yuan:2018inv,Chen:2016qju,Agaev:2020zag,Agaev:TJP} with significant experimental improvement over the past decade. These building blocks of matter are expected to give new information on the non-perturbative aspects of QCD theory. Also in contrast to the neutral states, charged states provide evidence for tetraquark exotic states since they cannot be counted as charmonium states.

The $ c\bar{c}s \bar{s}$ tetraquarks $  X(4140), X(4274), X(4500) $ and $ X(4700) $ were first reported by LHCb in 2016 \cite{Aaij:2016nsc} in the $J/\psi \phi $ invariant mass distribution. Very recently the LHCb collaboration has also detected two new states $ Z_{cs}(4000)^+ $ and $ Z_{cs}(4220)^+ $ with the quark content $ c\bar{c}u\bar{s} $, and two additional states corresponding to the $ c\bar{c}s \bar{s}$ called as $ X(4685)$ and $X(4630) $ in three-body decay $ B^+\rightarrow J/\psi \phi  K^+ $ decays with the following mass and decay width values. The analysis is based on the combined proton-proton $ (pp) $
collision data collected using the LHCb detector in Run 1 at centre-of-mass energies $ \sqrt {s} $ of 7 and $ 8 $ TeV, corresponding to a total integrated luminosity of $ 3 fb^{-1} $, and in Run 2 at
$ \sqrt {s}=13 $ TeV corresponding to an integrated luminosity of $ 6 fb^{-1} $. Among them $ X(4685) $ state decaying to the $J/\psi \phi  $ final state is observed with high significance. Now we'll call this particle $ X_{cs} $ for brevity:
\begin{eqnarray*}\label{Mass and width}
m_{X_{cs}}&=& 4684.3\pm 7^{+13}_{-16}~\mathrm{MeV},\\
\Gamma_{X_{cs}}&=&126\pm 15^{+37}_{-41}~\mathrm{MeV}
\end{eqnarray*}
and the quantum numbers is well determined. The BESIII collaboration \cite{Ablikim:2020hsk} reported the discovery of the first candidate for a charged hidden-charm tetraquark with strangeness, tentatively dubbed $ Z_{cs}(3985)^- $. It is unclear whether the new $ Z_{cs}(4000)^+ $ tetraquark can be identified with this state. Though their masses are consistent, the width of the BESIII particle is ten times smaller. The one LHCb observed is much broader, which would make it more natural to interpret as a compact tetraquark candidate.

This paper is structured as follows. In Section II, we briefly
mention about the QCD sum rules model and then determine sum
rules of the heavy-light $ X(4685) $ resonance in the tetraquark assumption. In Section III, the numerical analysis of sum rules are performed and
the results are compared with those of the other predictions
obtained in the literature. In the last section
we summarize our results. 
\section{Sum Rules Formalism}
Sum rules technique is a QCD based theoretical framework which
incorporates nonperturbative effects universally order by order and has some  advantages in exploring hadron characteristics including nonpertubative QCD, and have also been applied for the studies of the exotic hadrons. The QCD Sum rules is firstly proposed in studies \cite{Shifman:1978bx,Shifman:1978by,Ioffe:1981kw}  and also successfully applied to various baryonic and mesonic states \cite{Reinders:1984sr}.
According to the QCD sum rule approach, a two-point correlation
function is described as
\begin{equation}\label{eq:CorFunction}
\Pi_{\mu\nu}(p)=i\int d^{4}x~e^{ip\cdot x}\langle 0|
T\{J_{\mu}(x)J_{\nu}^{\dag}(0)\}|0\rangle,
\end{equation}
$J_{\mu}(x)$ is the interpolating current of the
related state and $T$ is the
time-ordering operator.

Correlation function can be computed in two different ways; it can
be expressed with regard to hadronic degrees of freedom such as the
masses, decay constants, form factors, the coupling constants and
so forth sandwiching complete sets of hadronic states with the same quantum numbers into the
correlation function. This side is called as "\textit{phenomenological}"
(or physical/hadronic) side. It can also be computed according to quark-gluon degrees of freedom in the deep Euclidean region i.e.,
when $q^2\rightarrow\infty$. That side is named as "\textit{theoretical}"
(or OPE/QCD) side. As usual we calculate the hadronic parameters
by equating the physical representation of the two point
function with the Operator Product Expansion (OPE) using Borel transformation in terms of momentum to
suppress the contributions coming from the higher states and continuum.\\\\
{\textbf{\textit{Phenomenological Side}}}:
The first stage of our computation is to obtain the phenomenological side
for the considered state. In order to apply the QCD sum rules
to investigate $ X(4685) $ meson, first we insert a complete set of states between the two currents in the \textit{T}-product definition in Eq.~(\ref{eq:CorFunction}). By
placing the complete set of intermediate states and performing integral over $x$, the phenomenological side
of the correlation function can be written in the following form:
\begin{eqnarray} \label{PhysSide}
\Pi_{\mu\nu}^{phen}(p)&=&\frac{\langle 0 \mid J_{\mu}\mid X_{cs}(p)\rangle
\langle X_{cs}(p) \mid J_{\nu}^{\dag} \mid 0 \rangle}
{(p^2-m_{X_{cs}}^2)}\notag\\
&+&\cdots~,
\end{eqnarray}
where "dots" stands for the contribution of higher states and
the continuum. To continue we define the pole residue as following matrix element:
\begin{eqnarray}\label{MatrixElements}
\langle 0|J_{\mu}|X_{cs}(p)\rangle=\lambda_{X_{cs}} \varepsilon_{\mu},
\end{eqnarray}
where $\varepsilon_{\mu}$ denotes the
polarization vector satisfying following relation:
\begin{eqnarray}\label{eq:polarizationt1}
\varepsilon_{\mu}\varepsilon^{*}_{\nu}=-g_{\mu\nu}+\frac{p_\mu p_\nu}{m_{X{{cs}}}^2}.
\end{eqnarray}
Below we present the correlation function belonging to the physical side 
in terms of the ground state mass and pole residue:
\begin{eqnarray}\label{eq:Corstructure}
\Pi_{\mu \nu }^{\mathrm{phen}}(p)&=&\frac{\lambda_{X_{cs}}^2}{{m_{X_{cs}}}^{2}-p^{2}} \bigg (-g_{\mu\nu}+\frac{p_\mu p_\nu} {m_{X_{cs}}^{2}} \bigg )\notag\\
&+& \ldots.
\end{eqnarray}
Here, we select the first structure, i.e. including $ g_{\mu\nu} $ terms.\\\\
%
{\textbf{\textit{Theoretical Side}}}:
To carry out the calculations, the second stage of the evaluation
is to define the theoretical side in the framework of QCD sum
rules writing the two-point correlation function firstly as in
Eq.~(\ref{eq:CorFunction}). The possible tetraquark interpolating current with $ J^P=1^+ $ for the $X_{cs}$ can be constructed as;
\begin{eqnarray} \label{current}
J_{\mu}^{X_{cs}}&=& i \epsilon_{abc} \epsilon_{dec} \Big( s^T_a C \gamma_5 c_b\Big) \Big(\bar{s}_d \gamma_{\mu} C \bar{c}^T_e\Big),
\end{eqnarray}
where $ a $ and $ b $ are color indices and $ C $ denotes the charge conjugation. 

Inserting the interpolating current in the correlation function in Eq.~(\ref{eq:CorFunction}) and employing Wick theorem after contractions, we obtain the theoretical (OPE) part of the correlation function expressed in terms of the light and heavy quark propagators. So after some algebra, the expression is obtained as follows:
\begin{eqnarray}\label{eq:contraction}
&&\Pi^{\mathrm{OPE}}_{\mu\nu}(p^{2})=i C C' \int d^{4}x~e^{ip \cdot x}\Big[Tr \Big(\gamma_{\mu}  \widetilde{S}_c^{e'e}(-x) \gamma_{\nu}\notag \\
&& \times S_s^{d'd}(-x)\Big) Tr \Big(\gamma_5 \widetilde{S}_s^{aa'}(x) \gamma_5 S_c^{bb'}(x)\Big)\Big], 
\end{eqnarray}
here 
\begin{eqnarray}
C=\varepsilon_{abc}\varepsilon_{dec},~~ C'=\varepsilon_{a'b'c'}\varepsilon_{d'e'c'},	
\end{eqnarray}
and we employed short-hand notation $ \widetilde{S}^{jj'}(x) = CS^{jj'T}(x)C $ in Eq.~(\ref{eq:contraction}).
In calculations, it is appropriate to employ the $x$-space
definition of the light quark propagators, while using
momentum-based expressions for the heavy quarks. The heavy quark
propagator's explicit form is \cite{Reinders:1984sr};
\begin{eqnarray}\label{HeavyProp}
&&S_{Q}^{ab}(x)=i\int \frac{d^{4}k}{(2\pi )^{4}}e^{-ik\cdot x}\Bigg
\{\frac{\delta_{ab}\left( {\slashed k}+m_{Q}\right) }{k^{2}-m_{Q}^{2}}    \notag \\
&&-\frac{g_{}G_{ab}^{\alpha \beta }}{4}\frac{\sigma _{\alpha \beta }\left( {%
\slashed k}+m_{Q}\right) +\left( {\slashed k}+m_{Q}\right) \sigma
_{\alpha\beta }}{(k^{2}-m_{Q}^{2})^{2}}   \notag \\
&&+\frac{g_{}^{2}G^{2}}{12}\delta _{ab}m_{Q}\frac{k^{2}+m_{Q}{\slashed k}}{%
(k^{2}-m_{Q}^{2})^{4}}+\ldots %
\Bigg \}.
\end{eqnarray}%
Here
\begin{eqnarray}\label{GluonField}
&&G_{ab}^{\alpha \beta }=G_{A}^{\alpha \beta
}t_{ab}^{A},\,\,~~G^{2}=G_{\alpha \beta }^{A}G_{\alpha \beta
}^{A},
\end{eqnarray}
where $a,\,b=1,2,3$ are the color index and $A,B,C=1,\,2\,\ldots
8$ are flavor indices, $t^{A}=\lambda ^{A}/2$, $\lambda ^{A}$ are
the Gell-Mann matrices and the gluon field strength tensor
$G_{\alpha \beta}^{A}\equiv G_{\alpha \beta }^{A}(0)$ is fixed at
$x=0$.
The light quark propagator is defined as:
\begin{eqnarray}\label{LightProp}
&&S_{q}^{ab}(x)=i\delta _{ab}\frac{\slashed x}{2\pi ^{2}x^{4}}-\delta _{ab}%
\frac{m_{q}}{4\pi ^{2}x^{2}}-\delta _{ab}\frac{\langle
\overline{q}q\rangle
}{12}  \notag \\
&&+i\delta _{ab}\frac{\slashed xm_{q}\langle \overline{q}q\rangle }{48}%
-\delta _{ab}\frac{x^{2}}{192}\langle \overline{q}g_{}\sigma
Gq\rangle
+i\delta _{ab}\frac{x^{2}\slashed xm_{q}}{1152}\langle \overline{q}%
g_{}\sigma Gq\rangle  \notag \\
&&-i\frac{g_{}G_{ab}^{\alpha \beta }}{32\pi ^{2}x^{2}}\left[ \slashed x{%
\sigma _{\alpha \beta }+\sigma _{\alpha \beta }}\slashed x\right]
-i\delta _{ab}\frac{x^{2}\slashed xg_{}^{2}\langle
\overline{q}q\rangle ^{2}}{7776} \notag \\
&&-\delta _{ab}\frac{x^{4}\langle \overline{q}q\rangle \langle
g_{}^{2}G^{2}\rangle }{27648}+\ldots.
\end{eqnarray}

Then taking into account tensor structure of $\Pi_{\mu\nu}^{\mathrm{OPE}}(p)$
we can write: 
\begin{eqnarray}
	\Pi_{\mu\nu}^{\mathrm{OPE}}(p) & = & \Pi_{0}^{\mathrm{OPE}}(p^{2})\frac{p_{\mu}p_{\nu}}{p^{2}} \notag\\ &+&\Pi_{1}^{\mathrm{OPE}}(p^{2})(-g_{\mu\nu}+\frac{p_{\mu}p_{\nu}}{p^{2}}),
\end{eqnarray}
here $ \Pi_{0}^{\mathrm{OPE}}(p^{2}) $ and $\Pi_{1}^{\mathrm{OPE}}(p^{2}) $ are invariant functions. The sum rules for the mass and pole residue of $X_{cs}$ and its \textit{b}-partner can be extracted
after equating the same structures in both $\Pi
_{\mu\nu}^{\mathrm{\textit{phen}.} }(p)$ and $\Pi_{\mu\nu}^{\mathrm{OPE}}(p)$. To
continue our evaluations, we select the same structures for each
one at the later stage. The invariant function $\Pi
^{\mathrm{OPE}}(p^{2})$ corresponding to this structure can be
represented as the dispersion integral
\begin{equation}\label{PiOPE}
\Pi^{\mathrm{OPE}}(p^{2})=\int_{\mathcal{M}^2}^{\infty}ds~\frac{\rho
^{\mathrm{OPE}}(s)}{s-p^{2}},
\end{equation}
where $ \mathcal{M}^2=4(m_c+m_s)^2 $ and $\rho^{\mathrm{OPE}}(s)$ is
the two-point spectral density. It includes terms with two
different contents and can be classified as
\begin{eqnarray} \label{Rho}
\rho^{\mathrm{OPE}}(s)&=&\rho
^{Pert.}(s)+\rho^{<qq>}(s)+\rho^{<GG>}(s)\notag\\ &+&\rho^{<qGq>}(s)+\rho^{<\bar{q}q>^2}(s)+\rho^{\langle \overline{q}q\rangle \langle g_{}^{2}G^{2}\rangle},~~~~
\end{eqnarray}
here the first term denotes perturbative contribution and the other terms denote non- perturbative contributions. 
The imaginary parts of the $\Pi_{n}$ functions leads us the
spectral densities $\rho_n(s)$, i.e.
$\rho_n(s)=\frac{1}{\pi}Im[\Pi_{n}]$. Due to the lengthy
expressions, only the perturbative part
$\rho^{Pert.}(s)$ of the spectral function is
given here:
\begin{eqnarray}
&& \rho^{Pert.}(s)= \int_{0}^{1}\int_{1}^{1-x_1}
dx_1 dx_2\frac{(\alpha \eta s- m_c^2 \beta \zeta)^2}{3072 \alpha \pi^6 \beta^8}\notag\\
&&\times 
\Big[35 \alpha^2 \eta^3 s^2 - 
2 \alpha m_c s \beta \Big(6 \eta m_s \beta (5 x_1+ 4 x_2) \notag\\
&&  + 13 \eta^2 m_c \zeta \Big)+ 3 m_c^2 \beta\Big(4 m_c m_s \beta^2 (2 x_1 + x_2) \zeta\notag\\ 
&& +\eta m_c^2 \beta \zeta^2+ 24 m_s^2 (x_2^2 + \sigma \zeta)^3 \Big)\Big], 
\end{eqnarray}
here
\begin{eqnarray*}\label{coeff}
x_1 + x_2-1= \alpha, \notag\\ 
x_2-1=\beta,\notag\\ 
x_1-1=\sigma, \notag\\ 
x_1 + x_2=\zeta,\notag\\ 
x_1 x_2=\eta. 
\end{eqnarray*}
After applying the Borel transformation on the variable $p^{2}$ to
both the phenomenological and OPE sides of the equality, subtracting
the contribution of higher resonances and continuum states and
assuming the quark-hadron duality, we find the required sum rules.

In the Borel scheme, the action of the Borelization
operator is given by 
$ \widehat{\textbf{B}}^2(p^2\rightarrow \tau) \{1/(s-p^2)\} =e^{-s\tau}$ and the OPE side of the correlation function can
be composed as perturbative and non-perturbative parts:
\begin{eqnarray} \label{BorelOPE}
\widehat{\textbf{B}}\Pi^{OPE}(p^2)&=&\widehat{\textbf{B}}\Pi^{Pert.}(p^2)+\widehat{\textbf{B}}\Pi^{\langle \overline{q}q \rangle}(p^2)\nonumber \\
&+&\widehat{\textbf{B}}\Pi^{ \langle GG \rangle}(p^2)+\widehat{\textbf{B}}\Pi^{\langle qGq \rangle}(p^2)\nonumber \\
&+&\widehat{\textbf{B}}\Pi^{<\bar{q}q>^2}(p^2)+\widehat{\textbf{B}}\Pi^{\langle \overline{q}q\rangle \langle g_{}^{2}G^{2}\rangle}(p^2)~~~
\end{eqnarray}
where $Pert.$ denotes the perturbative part and the upper indices
$ \langle \overline{q}q \rangle $,  $ \langle GG \rangle $,  $ \langle qGq \rangle$, $ \langle\bar{q}q \rangle ^2 $  and $ \langle \overline{q}q\rangle \langle g^{2}G^{2}\rangle $ represent the contributions of
quark, gluon and mixed condensates, respectively. 

Now using these definitions, transferring the continuum contribution to the QCD part, applying Borel transformation to both parts of the sum rules and equating them,  pole residue sum rule for the axial-vector meson $X_{cs}$ up to the dimension-seven condensates is written as follows:
\begin{eqnarray}\label{eq:lamdaSR}
\lambda_{X_{cs}}^{2}e^{-m_{X_{cs}}^{2}/M^{2}}=\int_{\mathcal{M}^2}^{s_{0}}ds\rho^{\mathrm{OPE}}(s)e^{-s/M^{2}}
\end{eqnarray}
and then taking the derivative of Eq.~(\ref{eq:lamdaSR}) in terms of $(-1/M^2)$ we reach the  mass sum rule of $ X_{cs} $:
\begin{equation}\label{eq:massSR}
m_{X_{cs}}^{2}=\frac{\int_{\mathcal{M}^2}^{s_{0}}dss\rho^{\mathrm{OPE}}(s)e^{-s/M^{2}}}{\int_{\mathcal{M}^2}^{s_{0}}ds\rho^{\mathrm{OPE}}(s)e^{-s/M^{2}}},
\end{equation}
where $s_0$ is the effective threshold for the onset of higher states and $ M $ is the auxiliary Borel parameter. The next step is to carry out the numerical analysis to determine the values of hadronic parameters of resonance $ X_{cs} $ and also replace $ c $ quark with $b$ quark to obtain the $b$-partner $ X_{bs} $ of $ X_{cs} $  in tetraquark picture.

Lets perform numerical analysis to complete the computation
considering $X_{cs}$ and its \textit{b}-partner in diquark-antidiquark picture.

\section{Mass and pole residue analysis}

After getting mass and pole residue sum rules as in Eq.~(\ref{eq:lamdaSR}) and  Eq.~(\ref{eq:massSR}), we will perform the numerical analyses to calculate observable characteristics of the hadronic ground state. We first give all the input values that are relevant for our
calculation in this section. For the quark masses, we employ $m_{s}=93^{+11}_{-5}\mathrm{MeV} $, $ m_c=(1.27\pm0.02)~\mathrm{GeV}$ and $m_b=(4.18^{+0.03}_{-0.02})~\mathrm{GeV} $ \cite{Zyla},  
$m^2_0=(0.8\pm0.2)~\mathrm{GeV^2}$
\cite{Dosch:1988vv,Belyaev:1982cd}, $\langle \bar{u}u\rangle$ or
$\langle \bar{d}d\rangle=-(0.24\pm0.01)^3~\mathrm{GeV^3}$, $\langle  0| \bar{s}s  |0\rangle =m_0^2\langle \bar{u}u \rangle$ \cite{Shifman:1978bx,Reinders:1984sr}
 and $\langle\frac{\alpha_sG^2}{\pi}\rangle =0.012~\mathrm{GeV}^4$ ~\cite{Ioffe:2005ym}.

\begin{figure}[h!]
	\centering
	\includegraphics[width=9cm]{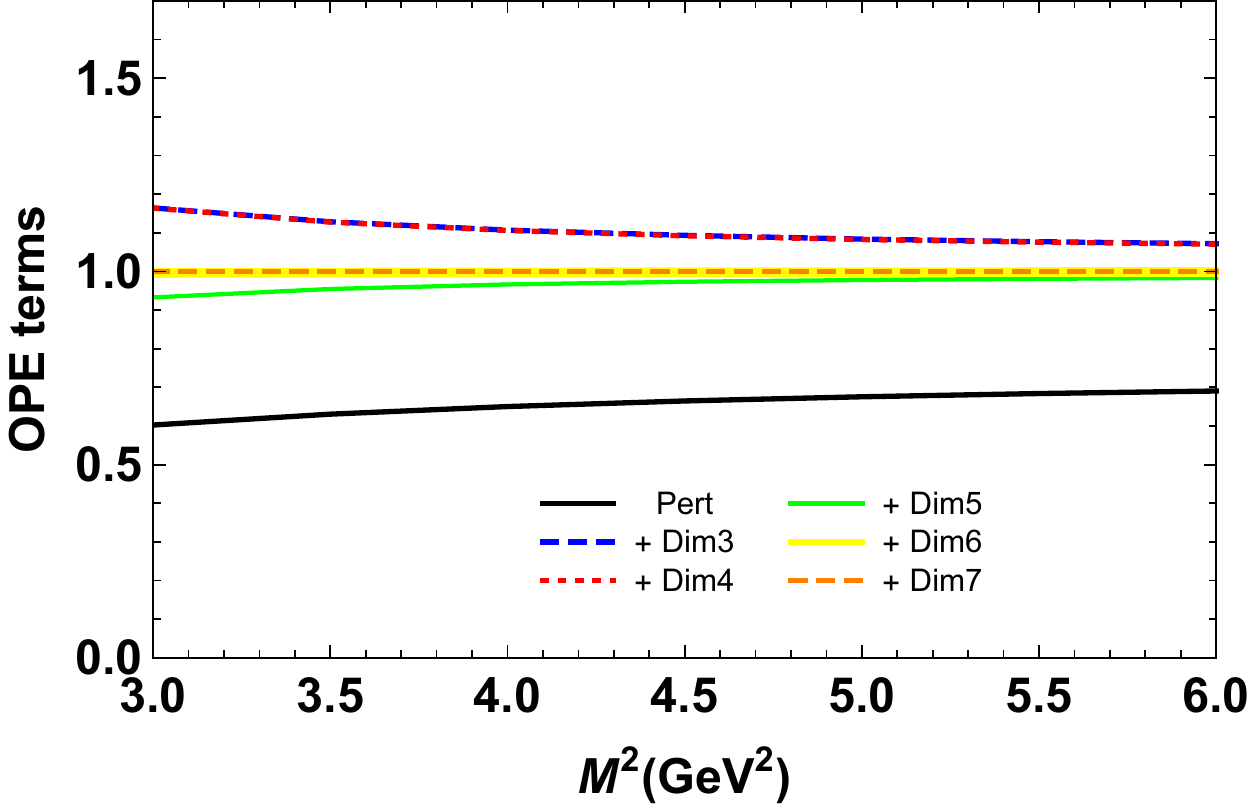}
	\caption{The OPE convergence of the sum rules for the $X_{cs}$ state: the ratio of the sum of the contributions up to specified dimension to the total contribution
		is drawned with respect to $ M^2 $ at fixed $ s_0=26.8 ~\mathrm{GeV^2} $ in the tetraquark picture.} 
	\label{OPE}
\end{figure}

\begin{figure}[h!]
	\centering
	\includegraphics[width=9cm]{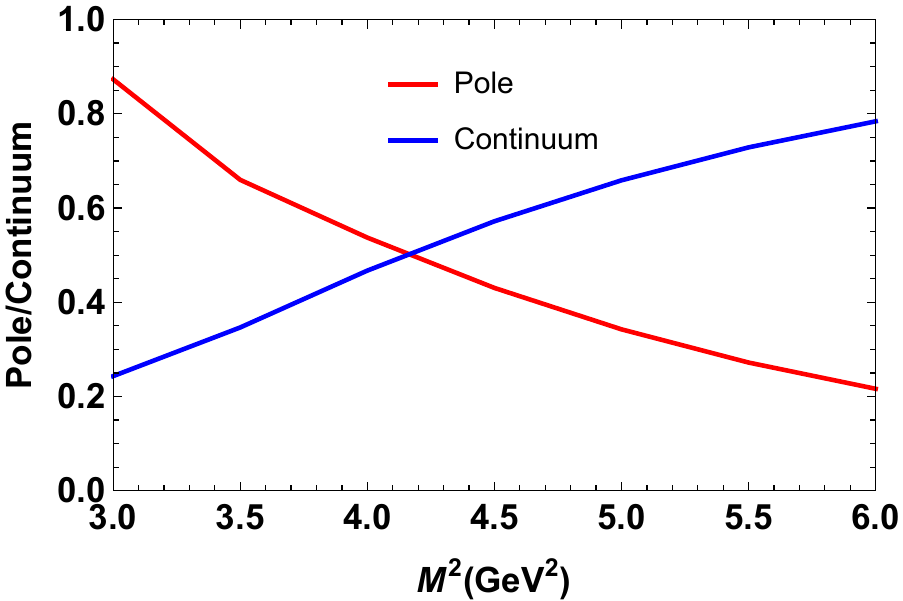}
	\caption{Pole and continuum contribution of the $ J^{PC} = 1^{+} $ hidden-charm state $X_{cs}$ at fixed $ s_0=26.8 ~\mathrm{GeV}^2$ in the tetraquark picture.} 
	\label{PoleCont}
\end{figure}

According to the idea of the sum rule the Borel mass $ M $ is an unphysical parameter, which is not related with the ground state mass. Therefore we have to show that resonance mass curve should be stable with changing $ M $ values. This is a criterion for the reliability of obtained sum rules. However it is not enough for us, we must provide another two norm that always need to be tested to guarantee the validity of the sum rules: \\\\
$ \textbf{-} $ \textit{The first one} is the lower bound of the
Borel window which is fixed by the convergence of the OPE. But, it is very hard to calculate up to high orders due to our lack of knowledge on the high-dimensional
condensates, so it is not possible to define a severe convergence standard. Instead, we use the contribution of the highest dimensional term
is less than $ 20\% $ of all the OPE terms (see Fig.~\ref{OPE}): 
\begin{eqnarray*}\label{eq:convergence}
\frac{\mathrm{\Pi}^\mathrm{Dim7}}{\mathrm{\Pi}^{\rm all~terms}_{OPE}}<0.2
\end{eqnarray*}\\
$ \textbf{-} $ \textit{The latter one} is the upper bound of the Borel window which is determined from the relative contribution of
the pole terms to the continuum as presented in Fig.~\ref{PoleCont}. The most frequently used condition is:
\begin{eqnarray*}\label{eq:PC}
\frac{\mathrm{\Pi}(s_0,M^2)}{\mathrm{\Pi}(\infty,M^2 )}\geq 0.5.
\end{eqnarray*}
\\
\\
Also, we obtain the best result for the continuum threshold  using $(m+0.5~\mathrm{GeV}) \leq \sqrt{s_0}\leq (m+0.55~\mathrm{GeV})$. After analysis we get the following working regions for the $X_{cs}$:
\begin{eqnarray*}\label{Msq and so value for X_{cs}}
&&M^2 \in [4.4-4.8] ~\mathrm{GeV^2}, \\
&&~~s_0 \in [26.8-27.5] ~\mathrm{GeV^2}.
\end{eqnarray*}

Finally we present our results in Table ~\ref{mresults} and ~\ref{fresults}:
\begin{table}[h!]
\caption{Estimations for the masses of $X_{cs}$ and $X_{bs}$ resonance.}\label{mresults}
\begin{tabular}{c||c|c}
\hline
                        		&$m_{{X}_{cs}}~\mathrm{(MeV)}$ &$m_{{X}_{bs}}~\mathrm{(MeV)}$  		\\
\hline\hline
Our results             		& $ 4607^{+36}_{-22} $ &$10604-10924$ \\
Experiment \cite{Zyla}          & $ 4684.3\pm 7^{+13}_{-16} $ & $ - $ \\
\hline	
\end{tabular}
\end{table}
\begin{table}[h!]
 \caption{Predictions for the pole residues of $X_{cs}$ and $X_{bs}$ state.}\label{fresults}
\begin{tabular}{c||c|c}
\hline
								& $\lambda_{X_{cs}}~\mathrm{(GeV^5)}$ &  $\lambda_{X_{bs}}~\mathrm{(GeV^5)}$   \\
\hline\hline
Our results             		& $6.19^{+0.32}_{-0.24}\times 10^{-2}$  	&$(42.2-53.7)\times 10^{-2}$ 						\\
Experiment \cite{Zyla}          & $  -  $    	& $ -$                         \\
\hline
\end{tabular}
\end{table} \\ 
\begin{figure}[h!]
\begin{center}
\includegraphics[width=9cm]{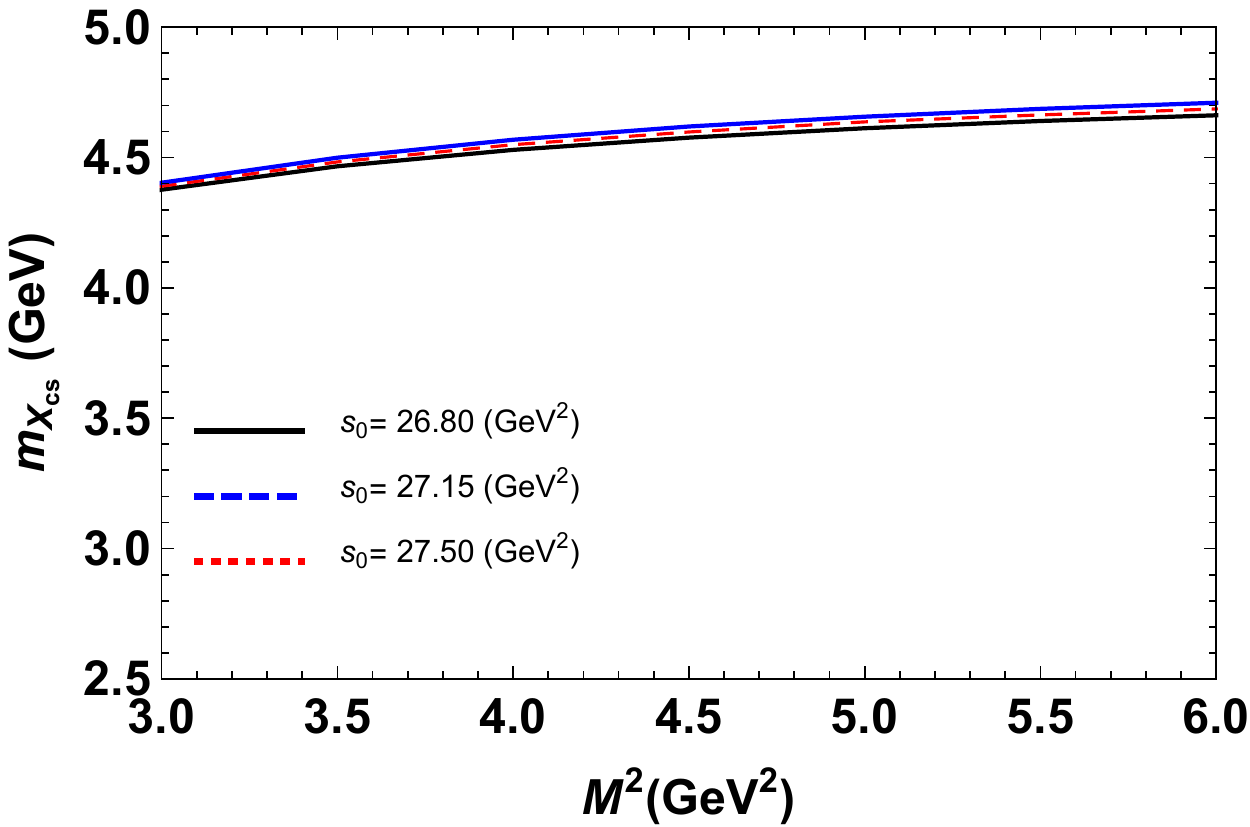}
\end{center}
\caption{The vacuum mass of the $X_{cs}$ state versus Borel mass parameter $ M^2 $  in the tetraquark picture for fixed different values of $s_0$.} \label{pl:massM2}
\end{figure}
\begin{figure}[h!]
\begin{center}
\includegraphics[width=9cm]{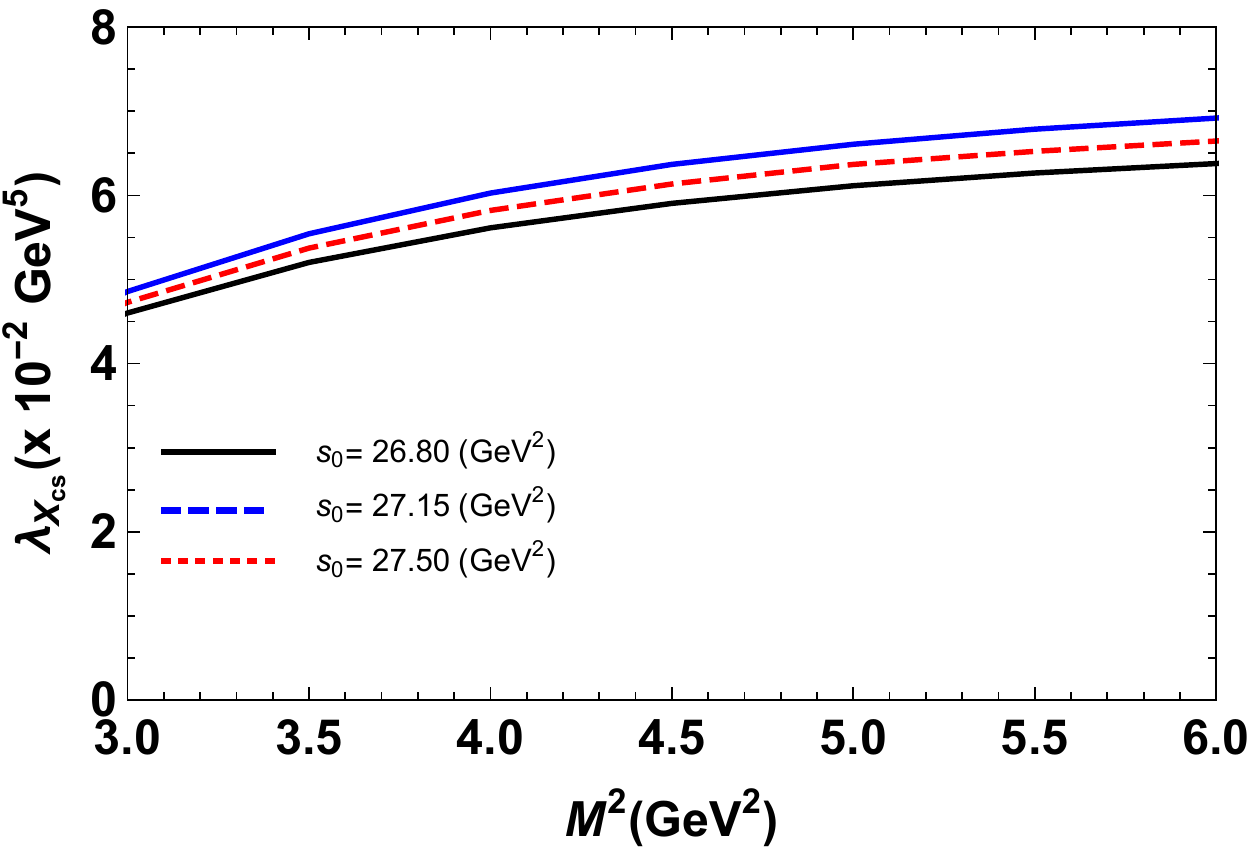}
\end{center}
\caption{The pole residue of the $X_{cs}$ resonance in terms of Borel mass parameter $ M^2 $ in the tetraquark picture for fixed different values of $s_0$.} \label{pl:fM2}
\end{figure}

Further to assure the stability region we plot the resonance mass and pole residue versus $ M^2 $ in Figs.~\ref{pl:massM2} and ~\ref{pl:fM2}. In Ref.~\cite{Wang:2021ghk}  $X_{cs}$ is analysed as the first excited state of $ X(4140) $ in tetraquark scenario and obtained the mass as $ M_{X}=4.70\pm0.12~\mathrm{GeV} $ and the pole residue $ \lambda_{X}=(1.08 \pm 0.17)\times 10^{(-1)} ~\mathrm{GeV^5}$ assigning the quantum numbers $J^{PC}=1^{++}$.
\section{Summary}
Hadron spectroscopy continues to be a rich area of fundamental exploration today, with results from collider experiments over the past two decades revealing the existence of multi-quark states more exotic than the familiar mesons and baryons. The emergence of exotic hadrons in experiments has provided new challenges for QCD. Among them  tightly bound colored diquark plays a fundamental role in hadron spectroscopy. Thanks to the remarkable achievements of LHC, a great amount of data on hadrons is accumulated. The first LHCb upgrade is currently in progress and data taking will recommence at the beginning of LHC Run 3 in 2022, with a second upgrade phase planned to gather a much larger data set by 2030. New tetraquark discoveries will have triggered the debate on the multi-quark states and allowed us to complete the hadron spectrum.

Testing QCD at high precision is a key to refine our understanding of strongly interacting matter, especially to explain the nature of tetraquark binding mechanisms. In this study, we analyse the very recently discovered $ X(4685) $ state and its \textit{b}-partner using the QCD Sum rules method including operators up to dimension seven. Within the error of uncertainties, our result $ m=4607^{+36}_{-22}~\mathrm{MeV}$ falls in the experimental measurement, that is $ 4684.3\pm 7^{+13}_{-16}~\mathrm{MeV} $. Additionally we search for the \textit{b}-partner of in the range of $ 10604~\mathrm{MeV} \leq m_{X_{bs}}\leq 10924~\mathrm{MeV}$. For the  pole residues we obtain $ \lambda_{X_{cs}}= 6.19^{+0.32}_{-0.24}\times 10^{-2}~\mathrm{(GeV^5)}$, $ \lambda_{X_{bs}}= (42.2-53.7)\times 10^{-2}~\mathrm{(GeV^5)}$ which can be used to study
the electromagnetic, weak or strong decays of handled states
as an important input parameter. 

From this result, we conclude that the assignments $ J^P = 1^+ $ and $ c\bar{c}s \bar{s}$ for the quantum numbers and quark structure of this state works well. Moreover hypothetical resonance $X_{bs} $ can be detected in future experiments. We hope to see whether other facilities confirm the LHCb observations and provides a new horizon for our understanding of the exotic structures in quantum chromodynamics (QCD).

\end{document}